\documentclass[pra, twocolumn,showpacs,nofootinbibfloatfix,amsmath,amsfonts,amssymb]{revtex4-1}%
\usepackage{amsmath,amsfonts,amssymb,color}
\usepackage{amsthm}
\usepackage{leftidx}
\usepackage{graphicx}
\usepackage{xcolor}
\usepackage{dcolumn}
\usepackage{bm}
\usepackage{epstopdf}
\usepackage{epsfig}

\newcommand{\n}{\noindent}

\begin{document}
	\title{{  Simulation of} non-Abelian braiding in Majorana time crystals}
	\author{Raditya Weda Bomantara}
	\email{phyrwb@nus.edu.sg}
	\affiliation{%
		Department of Physics, National University of Singapore, Singapore 117543
	}
	\author{Jiangbin Gong}%
	\email{phygj@nus.edu.sg}
	\affiliation{%
		Department of Physics, National University of Singapore, Singapore 117543
	}
	\date{\today}
	
	
	\vspace{2cm}
	
	\begin{abstract}
		Discrete time crystals have attracted considerable theoretical and experimental studies but their potential applications have remained unexplored.  A particular type of discrete time crystals, termed ``Majorana time crystals", is found to emerge
in a periodically driven superconducting wire accommodating two different species of topological edge modes. It is further shown that one
can manipulate different Majorana edge modes separated in the time lattice, giving rise to an unforeseen scenario for topologically protected gate operations {  mimicking braiding}.  The proposed protocol can also generate a magic state that is important for universal quantum computation.  This study thus advances the quantum control in discrete time crystals and  reveals their great potential arising from their time-domain properties.
		
	\end{abstract}
	
	\maketitle
	
	\textit{Introduction.} The idea of time crystals was first coined by Frank Wilczek in 2012 \cite{Wilczek}. Despite the existence of a no-go theorem which prohibits time crystals to arise in the ground state or equilibrium systems \cite{no}, time crystals in periodically driven systems, named discrete time crystals (DTCs), have recently attracted considerable interests \cite{Sacha,the1,the2,the3,the4,the5,the6,the7}.  Two experimental realizations of DTCs have been reported \cite{exp1,exp2}.
	
	Here we explore the potential applications of DTCs as exotic phases of matter \cite{tcrev}.  Specifically, DTCs are exploited to perform topologically protected quantum computation \cite{tqc,tqc2}. To that end, one needs to first find a particular type of DTCs that can simulate non-Abelian, e.g. Ising \cite{tqc,tqc2,ising,ising2,ising3} or Fibonacci \cite{tqc,tqc2,fib,fib2} anyons.  Ising anyons can be described in the language of Majorana fermions in one dimensional (1D) superconducting chains \cite{Ivanov,Kit,mich}.
	
DTCs have recently been proposed in a periodically driven Ising spin chain \cite{the5}. As learned from the mapping between one-dimensional (1D) superconducting chains and static spin systems \cite{ising,Ychen,map,DLoss}, we expect the emergence of DTCs in a periodically driven Kitaev superconducting chain.  Indeed, there period-doubling DTCs are obtained using the quantum coherence between
two types of topologically protected Floquet Majorana edge modes \cite{opt,M3,kk3}.
Such DTCs are termed `Majorana time crystals' (MTCs) below.  Next, a scheme is proposed to physically simulate the non-Abelian braiding of a pair of Majoranas
\cite{Tjun1,Tjun2,Ychen,wire,Tjun3,Tjun4} at two different time lattice sites.    We also elucidate how our scheme can be used to generate a magic state, which is necessary to perform universal quantum computation \cite{magic,magic2,magic3,magic4,magic5}. These findings open up a new concept in {  simulating} the braiding of Majorana excitations and should stimulate future studies of the applications of DTCs.
	
	\textit{Majorana time crystals.} Consider a periodically driven system $H(t)$ of period $T$. For the first half of each period, $H(t)$ is a 1D Kitaev chain with Hamiltonian $H_1= \sum_j^{N-1} \left( -J_j c_{j+1}^\dagger c_j + \Delta_j c_{j+1}^\dagger c_j^\dagger+h.c.\right) +\mu_{1} \sum_j^N c_j^\dagger c_j$
\cite{Kit}, and for the second half of each period,
$H(t)=	H_2 =\mu_{2} \sum_j^N c_j^\dagger c_j$. Here $c_j$ ($c_j^\dagger$) is the annihilation (creation) operator at site $j$, $J_j$ and $\Delta_j$ are respectively the hopping and pairing strength between site $j$ and $j+1$, $\mu_1$ and $\mu_2$ are chemical potential at different time steps. Throughout we work in a unit system with $\hbar=1$ .
Unless otherwise specified later,  we take $J_j=J$ and $\Delta_j=\Delta$ for all $j=1,\cdots N-1$ for our general discussions.
For later use, we also define the one-period propagator $\mathcal{U}=\mathcal{T} \exp\left(-\int_0^{T} \mathrm{i} H(t') dt'\right)$, where $\mathcal{T}$ is the time ordering operator.
One candidate for $H_1$ is an ultracold atom system \cite{opt,M3}, realizable by optically trapping 1D fermions inside a three dimensional (3D) molecular Bose-Einstein condensate (BEC). In such an optical lattice setup, the hopping term is already present due to the two Raman lasers generating the optical lattice, while the pairing term can be induced by introducing a radio frequency (rf) field coupling the fermions with Feshbach molecules from the surrounding BEC reservoir. Realizing the periodic quenching between $H_1$ and $H_2$ is also possible \cite{opt,M3}.
	
Our motivation for considering the above model system depicted by $H(t)$ is as follows. If $J=\Delta=\Delta^*$ and $\mu_1=0$, then $H(t)$ can be mapped to a periodically driven Ising spin chain \cite{Supp}, which is known to exhibit DTCs \cite{the5}.   We thus expect $H(t)$ to support DTCs.  That is, there exists some observable such that for a class of initial states, the oscillation in the expectation value of this observable does not share the period of $H(t)$, but exhibits a period of $nT$, with $n>1$ being stable against small variations in the system parameters. Furthermore, in the thermodynamic limit,  the oscillation of this observable with period $nT$ persists over an infinitely long time.

 DTCs in our model emerge from the interplay of periodic driving, hopping, and $p$-wave pairing. In particular,  $H(t)$ yields a number of interesting Floquet topological phases manifested by a varying number of edge modes, with their corresponding eigenphases of $\mathcal{U}$ being $0$ or $\pi$.   These  eigenmodes of $\mathcal{U}$ localized at the system edge are often called Floquet zero \cite{opt,kk3,M3} or $\pi$ edge modes \cite{opt,kk3,Derek1,Zhou1,R1,kk1,kk2}, possessing all the essential features of a Majorana excitation \cite{Supp}.  For example, by taking $\mu_2 T =JT=\Delta T= \pi$ and $\mu_1=0$, the eigenphases of $\mathcal{U}$ can be explicitly solved, which yield both Majorana zero and $\pi$ modes. Given that the Majorana zero ($\pi$) mode develops an additional phase $0$ ($\pi$) after one driving period $T$,
a superposition of Majorana zero and $\pi$ modes will evolve as a superposition, but with their relative phase being $\pi$ ($0$) after odd (even) multiples of $T$.  That is,
the ensuing dynamics yields period-doubling oscillations for a generic observable.  Further, because these edge modes are protected by the underlying topological phase, they do not rely on any fine tuning of the system parameters \cite{Supp}, yielding the necessary robustness for DTCs.

\begin{figure}
		\begin{center}
			\includegraphics[scale=0.32]{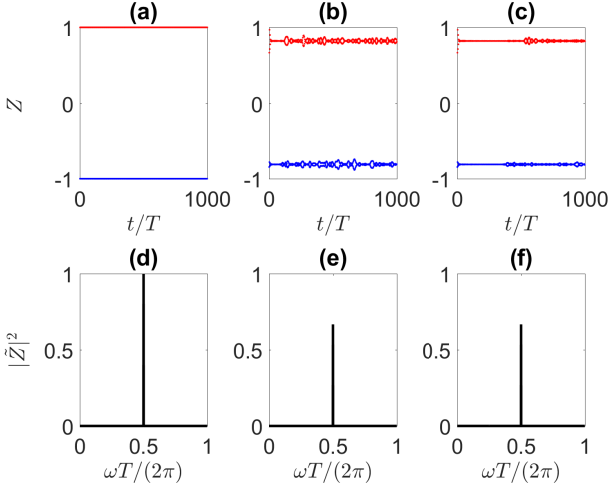}%
		\end{center}
		\caption{(color online). (a)-(c) Stroboscopic time evolution of $ Z$ evaluated at even (red) and odd (blue) integer multiples of $T$, {  given that $\Psi(0)=\gamma_1^A$}. The system parameters are (a) $\mu_1 T=0$, $J T=\Delta T=\mu_2 T=\pi$, $N=50$, (b) $\mu_1 T= 0.1 $, $\mu_2 T=3$, $\Delta T=1.5 J T=4.2 $, $N=50$, (c) same as (b) but with $N=200$. (d)-(e) Power spectrum associated with (a)-(c) shows a clear subharmonic peak at $\omega=\frac{\pi}{T}$.} 
		\label{tcpic}
	\end{figure}


Define two Majorana operators $\gamma_j^A=c_j+c_j^\dagger$ and $\gamma_j^B=\mathrm{i}(c_j-c_j^\dagger)$ at each chain site $j$, with $\gamma_j^A=(\gamma_j^A)^{\dagger}$, $(\gamma_j^A)^2=1$ and similar equalities for $\gamma_j^B$, as well as commutation relations $\{\gamma^A_j,\gamma^B_l\}=2 \delta_{AB}\delta_{jl}$.
{  In particular, before the periodic driving is turned on, the choice of system parameters above yield a Majorana zero mode $\Psi(0)=\gamma_1^A$. Once the driving is turned on, $\Psi(0)$ becomes a linear superposition of Majorana zero and $\pi$ modes and will then evolve non-trivially in time. At time $t$, it can be written in general as $\Psi(t)=\sum_j\sum_{l=A,B} c_{j,l}(t) \gamma_j^l$, where $\sum_j\sum_{l=A,B} |c_{j,l}(t)|^2=1$. To demonstrate how DTCs can be observed in the system, special attention is paid to the quantity $Z(t)=|c_{1,A}(t)|^2-|c_{1,B}(t)|^2$, which measures the difference between the weight of $\gamma_1^A$ and $\gamma_1^B$ in $\Psi(t)$.} 

Figure~\ref{tcpic}(a)-(c) show $Z$ vs time in several cases, whereas Fig.~\ref{tcpic}(d)-(f) show the associated subharmonic peak in the power spectrum, defined as $\tilde{Z}(\omega)=\sum_n Z(t)  \exp\left(\mathrm{i} n \omega T \right)$. $|\tilde{Z}(\omega)|^2$ is seen to
 be pinned at $\omega=\frac{\pi}{T}$, confirming the emergence of period-doubling DTCs.
   {  Under the special system parameter values chosen above, $\Psi(0)$ comprises of an equal-weight superposition of Majorana zero and $\pi$ modes (shown below) and will therefore undergo period-doubling oscillations between two Majorana operators $\gamma_1^A$ and $\gamma_1^B$ as time progresses. As Figs.~\ref{tcpic}(b) and (c) show, tuning the values of $\mu_1$, $J$, $\Delta$, and $\mu_2$ away from these special values still yields the same period-doubling oscillations for a long time scale, accompanied by some beatings in the time dependence which diminishes as the system size increases. These results thus justify the term
 MTC to describe such DTCs.}

\begin{figure}
		\begin{center}
			\includegraphics[scale=0.26]{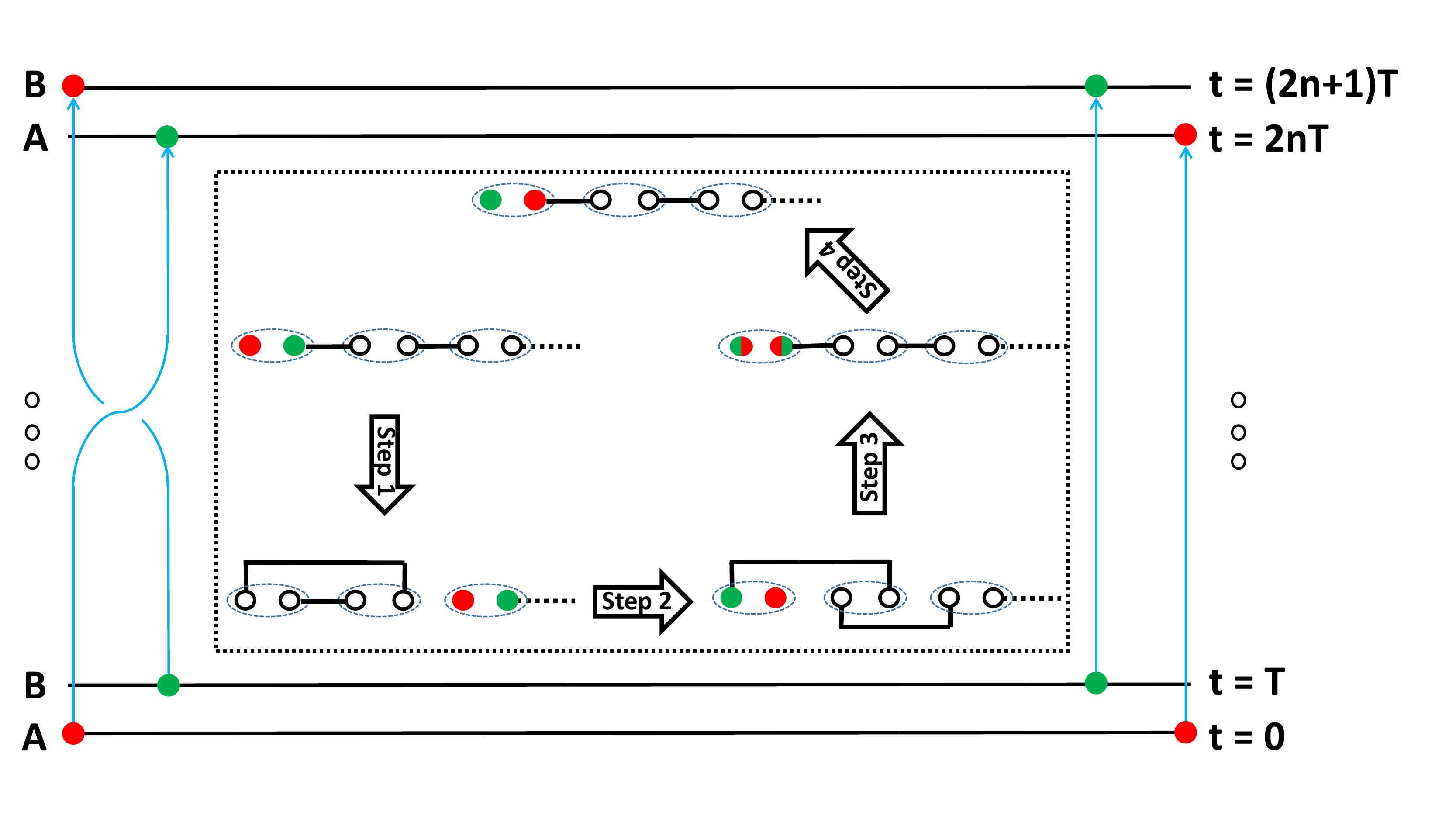}
		\end{center}
		\caption{(color online). Due to two nonequivalent time lattice sites labeled A and B, braiding of two left Majoranas separated in time {  can be simulated} by certain manipulations of the system. The inset shows the detail of our protocol. {  Red and green circles denote the Majorana modes localized at even and odd multiples of of the period} respectively, blue ellipses represent the lattice sites, {  empty circles denote the rest of the Majorana operators}, and black lines denote the coupling between two Majoranas due to $H_1$.}
		\label{tbpic}
	\end{figure}

	\textit{  Simulation of braiding protocol.} Consider now four Majoranas in our model, labeled as $\gamma_L^A$, $\gamma_R^A$, $\gamma_L^B$, and $\gamma_R^B$, with $\gamma_L^A$ ($\gamma_L^B$) and $\gamma_R^A$ ($\gamma_R^B$) representing Majorana edge modes localized in space, at the left and right edges respectively, and in time, at any even (odd) integer multiple of period. That is, $\gamma^B_L$ and $\gamma^B_R$ are obtained by evolving respectively $\gamma_L^A$ and $\gamma_R^A$ over one period.  During our protocol, $\gamma_L^A$ and $\gamma_L^B$ will be adiabatically manipulated {  to simulate braiding}, while $\gamma_{R}^{A}$ and $\gamma_R^B$ are left intact.  Such a nonconventional operation is schematically described by Fig.~\ref{tbpic}. Physical implementation of the adiabatic manipulation in the above-mentioned optical-lattice context {  \cite{opt}} can be done by slowly tuning the strength of the Raman lasers and the rf field.

\begin{figure}
		\begin{center}
			\includegraphics[scale=0.32]{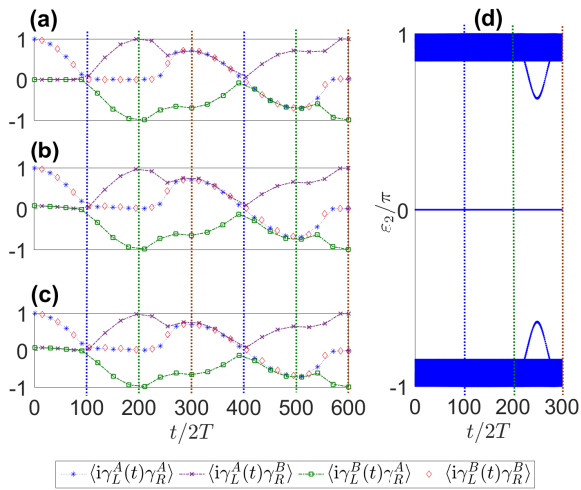}%
		\end{center}
		\caption{(color online) Time evolution of the Majorana correlation functions during the manipulation process for two different system parameters. Each step takes $200$ periods to complete. (a) $\mu_1 T=0$, $JT=\Delta T=\mu_2 T=\pi$, $N=100$. (b) $\mu_1 T= 0.3 $, $J T=3.3$, $\Delta T= 2.9 $, $\mu_2 T=3 $, $N=100$. (c) Same as (b) but in the presence of onsite, hopping, and pairing disorders, as well as small hopping term in $H_2$, averaged over $100$ disorder realizations. (d) Instantaneous eigenphases (denoted $\varepsilon_2$) of $\mathcal{U}^2$ during step 1-3, with $\varepsilon_2=0$ well separated from the bulk spectrum.  Blue, green, and brown vertical dotted lines mark the end of step 1, 2, and 3 respectively.}
		\label{bresult}
	\end{figure}

   Before presenting our protocol, we will first recast $H_1$ and $H_2$ in terms of Majorana operators as (focusing on the first three lattice sites {  and taking $\mu_1=0$})
 \begin{eqnarray}
 H_1&=& \mathrm{i} (\Delta_1/2-J_1/2)\gamma_1^A\gamma_2^B+ \mathrm{i}(\Delta_1/2+J_1/2)\gamma_1^B\gamma^A_2 \nonumber \\
 &+& \mathrm{i}\;\mathrm{Im}(\Delta_2/2-J_2/2)\gamma_2^A\gamma_3^A+ \mathrm{i}\;\mathrm{Re}(\Delta_2/2+J_2/2)\gamma_2^B\gamma^A_3  \nonumber \\
 &+& \mathrm{i}\;\mathrm{Im}(\Delta_2/2+J_2/2)\gamma_2^B\gamma_3^B+ \mathrm{i}\;\mathrm{Re}(\Delta_2/2-J_2/2)\gamma_2^A\gamma^B_3  \nonumber \\
 &+& \cdots \;,
 \label{meq}
  \end{eqnarray}
  and $H_2= -(\mu_2/2)(1+\mathrm{i} \gamma_1^A\gamma^B_1) +\cdots$, with $J_1$, $\Delta_1$, $J_2$, and $\Delta_2$ are subject to adiabatic manipulations, during which $J_2$ and $\Delta_2$ may be complex, while $J_j$ and $\Delta_j$ for $j\neq 2$ are assumed to be always real. For the sake of analytical solutions
  and better qualitative understandings, we again take $\mu_2 T=J_j T=\Delta_{j} T=\pi$ at the start to illustrate our idea, so that $\gamma_L^A$ ($\gamma_L^B$), initially prepared from the edge mode of $H_1$, is precisely $\gamma_1^A$ ($\gamma_1^B$). As demonstrated in Fig.~\ref{bresult}(a)-(c) below, this fine tuning of the system parameters is not needed in the actual implementation.

  In step 1, we exploit the adiabatic deformation of Majorana zero and $\pi$ modes, denoted $\hat{0}$ and $\hat{\pi}$, along an adiabatic path
     with $J_2=\Delta_2$ being real.
     To develop insights into this step,
    we parameterize $J_1+\Delta_1=2\pi/T$, $J_1-\Delta_1=2\pi\sin(\phi_1)/T$, $J_2=\Delta_2=2\pi\cos(\phi_1)/T$. As detailed in Supplementary Material \cite{Supp},
    we find (up to an arbitrary overall constant)
     \begin{eqnarray}
     \hat{0}=[\cos(\phi_1)\gamma_1^A-\sin(\phi_1)\gamma_3^A]+ [\cos(\phi_1)\gamma_1^B-\sin(\phi_1)\gamma_3^B]; \nonumber\\
     \hat{\pi}=[\cos(\phi_1)\gamma_1^A-\sin(\phi_1)\gamma_3^A]- [\cos(\phi_1)\gamma_1^B-\sin(\phi_1)\gamma_3^B]. \nonumber
     \end{eqnarray}

    By tuning $\phi_1$ slowly from $0$ to $\pi/2$, $\hat{0}$ will adiabatically
     change from $(\gamma_1^A+\gamma_1^B)$ to $-(\gamma_3^A+\gamma_3^B)$, whereas $\hat{\pi}$ will adiabatically
     change from $(\gamma_1^A-\gamma_1^B)$ to $(\gamma_3^B-\gamma_3^A)$, i.e., both zero and $\pi$ modes
     are now shifted to the third site. Due to this adiabatic following, a superposition of  $\hat{0}$ and $\hat{\pi}$ modes remains a superposition, thus preserving the DTC feature were the adiabatic process stopped at any time. The net outcome of this step can thus be described simply as $\gamma_1^A\rightarrow -\gamma_3^A$ and $\gamma_1^B\rightarrow -\gamma_3^B$.

     Step 2 continues to adiabatically deform $\hat{0}$ and $\hat{\pi}$.  Starting from $J_2=\Delta_2=0$ as a result of step 1, we consider an adiabatic path with $J_2=-\Delta_2$ being purely imaginary values. If we parameterize $J_1-\Delta_1=2\pi/T$, $J_1+\Delta_1=2\pi\cos(\phi_2)/T, J_2=-\Delta_2=\mathrm{i} \pi \sin(\phi_2)/T$, then one easily finds \cite{Supp}
     { 
      \begin{eqnarray}
     \hat{0}= [\sin(\phi_2)\gamma_1^B-\cos(\phi_2)\gamma_3^A]-[\sin(\phi_2)\gamma_1^A+\cos(\phi_2)\gamma_3^B]; \nonumber\\
     \hat{\pi}=[\sin(\phi_2)\gamma_1^B-\cos(\phi_2)\gamma_3^A]+[\sin(\phi_2)\gamma_1^A+\cos(\phi_2)\gamma_3^B].\nonumber
     \end{eqnarray}
    As $\phi_2$ adiabatically increases from $0$ to $\pi/2$,  $\hat{0}$ and $\hat{\pi}$ undergo further adiabatic changes to
     $(\gamma_1^B - \gamma_1^A)$ and $(\gamma_1^B +\gamma_1^A)$ respectively.} The overall transformation of this step is $-\gamma_3^A\rightarrow \gamma_1^B$ and $-\gamma_3^B \rightarrow -\gamma_1^A$.

    In step 3, we exploit further the coherence between Majorana zero and $\pi$ modes so as to recover the system's original Hamiltonian, while at the same time preventing $\gamma_L^A$ and $\gamma_L^B$ from completely untwisting and returning to their original configuration. As an innovative adiabatic protocol, we adiabatically change the system parameters {\it every other period}. This amounts to introduce a characteristic frequency $\pi/T$ in our adiabatic manipulation, {  resulting in the coupling between zero and $\pi$ quasienergy space. As the system parameters are adiabatically tuned, Majorana zero and $\pi$ modes will then adiabatically follow the degenerate eigenmodes of $\mathcal{U}^2$ (i.e.,
    	the two-period propagator) associated with zero eigenphase. This leads to a nontrivial rotation between the two Majorana modes dictated by the non-Abelian Berry phase in this degenerate subspace.  With this insight, one can envision many possible adiabatic paths to induce a desirable rotation between Majorana $0$ and $\pi$ modes. }


   After some trial and error attempts, we discover a class of adiabatic paths for step 3 that can yield a rotation of $\pi/4$ between $\gamma_L^A$ and $\gamma_L^B$. Specifically,  we fix $J_1$  and let {   $\Delta_1= 2\pi f_{3,a}(t)/T$, $J_2=\sqrt{2} \pi\exp\left(\frac{\mathrm{i} \pi}{4}\right)[1-\mathrm{i}f_{3,b}(t)]/T$, and $\Delta_2=\sqrt{2} \pi \exp\left(-\frac{\mathrm{i} \pi}{4}\right) [1+\mathrm{i}f_{3,c}(t)]/T$, where $f_{3,l}(t)$, with $l=a,b,c$, are certain (not necessarily the same) functions that slowly increase from -1 to 1 {\it for every other period}}.  That is, for each new period, $f_{3,l}(t)$ are alternatively increased or stay at the values of the previous step.
   At the end of the adiabatic manipulation, this step yields the original Hamiltonian, with $\gamma_1^B\rightarrow (\gamma_1^A+\gamma_1^B)/\sqrt{2}$ and $-\gamma_1^A\rightarrow (\gamma_1^B-\gamma_1^A)/\sqrt{2}$ to a high fidelity.

   Finally in step 4, we repeat the 3 steps outlined above to obtain the overall transformations $\gamma_L^A \rightarrow \gamma_L^B$ and $\gamma_L^B\rightarrow -\gamma_L^A$, which completes the simulated braiding operation to the two different species of Majoranas and at the same time resets the system configuration. As shown in Fig.~2, at the start of the protocol, $\gamma_L^A$ ($\gamma_L^B$) at our MTC appears at even (odd) multiples of $T$; by constrast, at the end of the protocol, $\gamma_L^A$ ($\gamma_L^B$) appears at odd (even) multiples of $T$.
  	
 To confirm the above analysis, we calculate the evolution of Majorana correlation functions during the manipulation process. The system is assumed to be in the even parity state such that initially $\langle \mathrm{i} \gamma_{L}^A \gamma_{R}^A\rangle = \langle \mathrm{i} \gamma_{L}^B \gamma_{R}^B\rangle = 1$ and $\langle \mathrm{i} \gamma_{L}^A \gamma_{R}^B\rangle = \langle \mathrm{i} \gamma_{L}^B \gamma_{R}^A\rangle=0$, where $\gamma_{i}^\alpha\equiv \gamma^{\alpha}_{i}(t=0)$, $\alpha=A,B$, and $i=L,R$. During the manipulation process, $\gamma_{A}^{L}(t)$ and $\gamma_B^L(t)$ in general become a superposition of $\gamma_{L}^{A}$ and $\gamma_{L}^{B}$, thus changing  the correlations $\langle \mathrm{i} \gamma^{\alpha}_L(t) \gamma^{\beta}_R\rangle$, where $\alpha,\beta=A,B$. The success of our protocol is then marked by the final correlation functions $\langle \mathrm{i}\gamma_{L}^A(t_f) \gamma_{R}^A\rangle =\langle \mathrm{i}\gamma_{L}^B(t_f) \gamma_{R}^B\rangle=0 $ and $\langle \mathrm{i}\gamma_{L}^A(t_f) \gamma_{R}^B\rangle =-\langle \mathrm{i}\gamma_{L}^B(t_f) \gamma_{R}^A\rangle=1 $. {  In experiment, Majorana correlation functions $\langle \mathrm{i} \gamma_L^A \gamma_R^A \rangle$ and $\langle \mathrm{i} \gamma_L^B \gamma_R^B \rangle$ may be measured via time-of-flight imaging method as proposed in Ref.~\cite{tof} or indirectly by measuring the parity of the wire \cite{read} at even and odd integer multiples of $T$. To measure cross correlation functions such as $\langle \mathrm{i} \gamma_L^B \gamma_R^A \rangle$, one could first turn off the periodic driving on the right half of the wire after the protocol is completed, then wait for one period. Since $\gamma_R^A$ is a Majorana zero mode in the absence of periodic driving by construction, it will stay invariant in one period, whereas the left Majorana mode will transform into $\gamma_L^B$ \cite{note}. The same readout process can then be carried out to measure their correlation functions.}
	
	The full evolution of Majorana correlation functions is depicted in Figs.~\ref{bresult}(a)-(c) under different system parameter values. In particular, Fig.~\ref{bresult}(c) assumes also the presence of disorders and a small hopping term in $H_2$, which may arise due to the presence of the Raman lasers, even after taking low frequency and large detuning values. More precisely, hopping, pairing, and onsite static disorders are considered by taking $J_j= J+\delta J_j$, $\Delta_j= \Delta +\delta \Delta_j$, $\mu_1\rightarrow \mu_1+\delta \mu_{1,j}$, and $\mu_2\rightarrow \mu_2+\delta \mu_{2,j}$, where $\delta J_j T$, $\delta \Delta_j T$, $\delta \mu_{1,j} T$ and $\delta \mu_{2,j} T$ uniformly take random values between $-0.1$ and $0.1$, while the small hopping term is of the form $-\sum_j \left(\mathcal{J}+\delta \mathcal{J}_{j}\right) c_{j+1}^\dagger c_j +h.c.$, where $\mathcal{J}T=0.025$ and $\delta \mathcal{J}_{j} T\in\left[-0.01,0.01\right]$. The fact that Figs.~\ref{bresult}(a)-(c) look qualitatively the same demonstrates the robustness of our protocol against such system imperfections. Finally, plotted in Fig.~\ref{bresult}(d) is the whole eigenphase spectrum of $\mathcal{U}^2$, which indicates that its zero eigenphases are well separated from the rest of the spectrum, thus confirming the topological protection needed to realize the rotation between $\gamma_L^A$ and $\gamma_L^B$.

	\textit{Discussion.}  {  Due to fermion parity conservation, a minimum of four Majoranas is required to harness their non-Abelian features for nontrivial (single-qubit) gate operations. Our work demonstrates, through exploiting the time-domain features, that this can be achieved in a minimal single wire setup, thus avoiding the necessity to design complicated geometries \cite{Tjun1,Tjun2,Tjun3,Tjun4}. Moreover, as demonstrated in Supplementary Material \cite{Supp}, our setup can be readily extended to an array of wires to simulate more intricate braiding between various pairs of Majoranas at different time and wires. In view of these two aspects, it is expected that certain quantum computational tasks may now be carried out using significantly less number of wires.}

	As another feature of the proposed protocol, the end of step 3 has achieved the transformation $\gamma_L^A\rightarrow \frac{1}{\sqrt{2}}\left(\gamma_L^A+\gamma_L^B\right)$ and $\gamma_L^B\rightarrow \frac{1}{\sqrt{2}}\left(\gamma_L^B-\gamma_L^B\right)$, which can be written as $V=\exp\left(-\frac{\pi}{8}\gamma_L^A\gamma_L^B\right)$. In the even parity subspace, a qubit can be encoded in the common eigenstates of the parity operators $\mathrm{i} \gamma_{L}^A\gamma_{R}^A $ and $\mathrm{i} \gamma_{L}^B\gamma_{R}^B $, such that $\mathrm{i} \gamma_{L}^A\gamma_{R}^A  |0\rangle =\mathrm{i} \gamma_{L}^B\gamma_{R}^B  |0\rangle=|0\rangle$ and $ \mathrm{i} \gamma_{L}^A\gamma_{R}^A |1\rangle =\mathrm{i} \gamma_{L}^B\gamma_{R}^B\rangle  |1\rangle =-|1\rangle$. It can be easily verified that $V$ maps $|0\rangle$ to a magic state $\cos\frac{\pi}{8}|0\rangle -\sin\frac{\pi}{8}|1\rangle$. It is known that a combination of Clifford gates and a magic state is required to achieve universal quantum computation \cite{magic,magic2}. While Clifford gates can be realized in a typical Ising anyonic model alone, the creation of a magic state normally requires an additional dynamical process \cite{magic3,magic4,magic5}.
{  The rather straightforward realization of the operation $V$ here is hence remarkable.}
	
	{  It is also important to ask to what extent the gate operations here share the robustness of the braiding of two Majorana modes.  On the one hand, if $J_2$, $\Delta_1$, and $\Delta_2$ in step 3 take arbitrary time-dependence then the desired braiding outcome cannot be achieved.
 On the other hand, to physically implement a braiding operation without the use of direct spatial interchange between two Majoranas,
  we expect some necessary control to restrict the time dependence of $J_2$, $\Delta_1$, and $\Delta_2$ used in step 3 to a certain degree.
  Our further investigations \cite{Supp} indicate that our protocol does enjoy some weak topological protection, in the sense that
  its fidelity is rather stable under considerable time-dependent deformation: $J_2\rightarrow J_2+\delta J_2$, $\Delta_1\rightarrow \Delta_1+\delta \Delta_1$, and $\Delta_2 \rightarrow \Delta_2 +\delta \Delta_2$ in step 3, where $\delta J_2$, $\delta \Delta_1$, and $\delta \Delta_2$ represent time-dependent perturbations, vanishing at the start and end of step 3, with relative strength of the order 5\% \cite{Supp}.}
	
	\textit{Conclusion.} A coherent superposition of Majorana zero and $\pi$ modes of a periodically driven 1D superconducting wire is shown to yield period-doubling MTCs. By adiabatic manipulation of the Majorana zero and $\pi$ modes, we have proposed a {  relatively} robust scheme to {  mimic the braiding of} two Majorana modes localized at different physical and time lattice sites. Our approach is promising for physical resource saving. As an important side result, we also obtain a magic state crucial for universal quantum computation \cite{magic,magic2}.

	\begin{acknowledgements}
\vspace{0.3cm}
	\n {\bf Acknowledgements:}  J.G. is supported by the Singapore NRF grant No. NRF-NRFI2017-04 (WBS No. R-144-000-378-281) and by the Singapore Ministry of Education Academic Research Fund Tier I (WBS No. R-144-000-353-112).
\end{acknowledgements}

	\onecolumngrid
	\appendix
	
	\vspace{1cm}
	
	\begin{center} {\bf Supplemental Material} \end{center}
	
	This supplemental material has eight sections. In Sec.~A, we elucidate in detail the exact mapping between $H(t)$ defined in the main text and a periodically driven Ising spin chain. This mapping allows us to deduce the parameter regime in which DTCs may emerge. {  In Sec.~B, we present some comparisons between our model and typical Kitaev Hamiltonian in static systems.} In Sec.~C, we formulate the definition of Majorana modes and Majorana time crystals in time periodic systems. In Sec.~D, the role of Majorana zero and $\pi$ modes in Majorana time crystals is discussed. In Sec.~E, we verify that $\hat{0}$ and $\hat{\pi}$ presented in the main text during the first two steps in the protocol are indeed Majorana zero and $\pi$ modes. {  In Sec.~F, we present some numerical results to demonstrate the robustness of step 3 in the protocol against small deformations in the adiabatic parameters}. {  In Sec.~G, we discuss the feasibility of our protocol in real experiment and possible error correction against decoherence}. Finally, we propose the possibility to extend our protocol to an array of superconducting wires in Sec.~H.

\section*{Section A: Exact mapping between a periodically driven superconducting wire and Ising spin chain}

Consider a periodically driven Ising spin chain as follows,

\begin{equation}
\mathcal{H}(t)=\begin{cases}
-\mathcal{J} \sum_j^{N-1} \sigma_{z,j+1}\sigma_{z,j} & \text{for } MT < t\leq (M+\frac{1}{2}) T \\
-f \sum_j^N  \sigma_{x,j} & \text{for }(M+\frac{1}{2})T<t<(M+1)T \\
\end{cases} \;, \label{ising}
\end{equation}

\n where $\sigma_{x(z),j}$ represent Pauli operators acting on site $j$, $J$ describes the strength of nearest neighbor spin-spin interaction, $M$ is an integer, $T$ is the period of the Hamiltonian, and $f$ corresponds to the strength of the Zeeman term. Following Ref.~\cite{map}, we define

\begin{eqnarray}
a_j&=& \prod_{l<j} \sigma_{x,l} \sigma_{z,j} \;, \\
b_j &=& \mathrm{i} a_j \sigma_{x,j} \;.
\end{eqnarray}

\n It can be easily verified that $a_j=a_j^\dagger$, $b_j=b_j^\dagger$, $\left\lbrace a_j,b_j\right\rbrace=0$, and $\left\lbrace a_j,a_k\right\rbrace=\left\lbrace b_j,b_k\right\rbrace=2\delta_{j,k}$. That is, $a_j$ and $b_j$ are Majorana operators. Upon expressing $\sigma_{x(z),j}$ in terms of $a_j$ and $b_j$,

\begin{eqnarray}
\sigma_{x,j} &=& -\mathrm{i} a_j b_j \;,\\
\sigma_{z,j+1}\sigma_{z,j} &=& -\mathrm{i} b_j a_{j+1} \;,
\end{eqnarray}

\n and Eq.~(\ref{ising}) becomes

\begin{equation}
\mathcal{H}(t)=\begin{cases}
\mathrm{i} \mathcal{J} \sum_j^{N-1} b_j a_{j+1} & \text{for } MT < t\leq (M+\frac{1}{2}) T\\
\mathrm{i} f \sum_j^N a_j b_j  & \text{for }(M+\frac{1}{2})T<t<(M+1)T \\
\end{cases} \;. \label{ising2}
\end{equation}

Given the Majorana operators $a_j$ and $b_j$, we can form complex fermionic operators as

\begin{eqnarray}
c_j &=& \frac{1}{2}\left(a_j -\mathrm{i} b_j\right) \;,\\
c_j^\dagger &=& \frac{1}{2}\left(a_j +\mathrm{i} b_j\right) \;,
\end{eqnarray}

\n so that

\begin{eqnarray}
\mathrm{i} a_j b_j &=& -\left(c_j^2+c_j^\dagger c_j -c_j c_j^\dagger +c_j^{\dagger 2}\right) \nonumber \\
&=& 2c_j^\dagger c_j-1 \;, \\
\mathrm{i} b_j a_{j+1} &=& -\left(c_jc_{j+1}+c_jc_{j+1}^\dagger-c_j^\dagger c_{j+1}-c_j^\dagger c_{j+1}^\dagger\right) \nonumber \\
&=& c_{j+1}^\dagger c_j -c_{j+1}^\dagger c_j^\dagger +h.c.
\end{eqnarray}

\n where we have used the fact that $c_j^2=c_j^{\dagger 2}=0$ and $c_j c_k^\dagger=\delta_{j,k}-c_k^\dagger c_j$. Eq.~(\ref{ising2}) can further be written as

\begin{equation}
\mathcal{H}(t)=\begin{cases}
\mathcal{J} \sum_j^{N-1} \left(c_{j+1}^\dagger c_j -c_{j+1}^\dagger c_j^\dagger  +h.c.\right) & \text{for } MT < t\leq (M+\frac{1}{2}) T\\
f \sum_j^N \left(2c_j^\dagger c_j -1\right)  & \text{for }(M+\frac{1}{2})T<t<(M+1)T \\
\end{cases} \;, \label{ising3}
\end{equation}

\n which is equivalent to $H(t)$ in the main text up to an unimportant constant upon identifying $J=\Delta=\Delta^*=-\mathcal{J}$, $\mu_2=2f$, and $\mu_1=0$.

{  \section*{Section B: Comparison with static Kitaev Hamiltonian}

We start by first writing $H(t)$ in momentum space as

\begin{eqnarray}
H(t) &=& \begin{cases}
\sum_k \frac{1}{2} \Psi_k^\dagger h_{1,k} \Psi_k & \text{for } MT < t\leq (M+\frac{1}{2}) T\\
\sum_k \frac{1}{2} \Psi_k^\dagger h_{2,k} \Psi_k & \text{for }(M+\frac{1}{2})T<t<(M+1)T \\
\end{cases} \;, \\
h_{1,k} &=& \mu_1 \tau_z - J \cos k \tau_z +\Delta \sin k \tau_y \;, \\
h_{2,k} &=& \mu_2 \tau_z \;,
\end{eqnarray}

\n where $\Psi_k =\left(c_k, c_{-k}^\dagger\right)^T$ is the Nambu vector, and $\tau_{a=x,y,z}$ are Pauli matrices acting on the Nambu space. In Nambu basis, the one-period propagator can be simply written as the product of two exponentials as (taking $T=2$ and $\hbar=1$ units)

\begin{equation}
U(k) = \exp\left(-\mathrm{i} h_{1,k}\right) \exp\left(-\mathrm{i} h_{2,k}\right) \;.
\end{equation}

\n In particular, since both $h_{1,k}$ and $h_{2,k}$ are both $2\times 2$ matrices, we may combine the two exponentials to write $U(k)=\exp\left(-\mathrm{i} h_f(k)\right)$, where $h_f(k)=\theta \hat{m}\cdot \tau$ is the Floquet Hamiltonian, where

\begin{eqnarray}
\theta &=& \arccos\left[ \cos (\mu_2) \cos(h) -\sin(\mu_2)\sin(h) \frac{\mu_1-J\cos(k)}{h} \right]\;, \\
\hat{m} &=& \frac{\Delta \sin(k)\sin(h)}{h\sin\theta} \left[\cos(\mu_2) \hat{y}-\sin(\mu_2) \hat{x} \right] +\frac{h\sin(\mu_2) \cos(h)+(\mu_1-J\cos(k))\cos(\mu_2) \sin(h)}{h\sin(\theta)}\hat{z} \;, \\
h &=& \sqrt{[\mu_1-J\cos(k)]^2+[\Delta \sin(k)]^2} \;.
\end{eqnarray}

Note that while $h_f$ takes the same structure as a typical Kitaev Hamiltonian in momentum space \cite{Kit}, the effective chemical potential, hopping strength, and (now complex) $p$-wave pairing, acquire non-trivial $k$ dependence. Moreover, since $h_f$ appears only as a phase, its eigenvalues (quasienergies) can only be defined up to a modulus of $2\pi$, which leads to the existence of two gaps at quasienergy zero and $\pi$. By contrast, Kitaev Hamiltonian in static systems has only one gap at energy zero. Despite these differences, one important similarity between $h_f$ and Kitaev Hamiltonian is the presence of particle hole symmetry, which for the former can be written explicitly by the operator $\mathcal{P}=\tau_x K$ which satisfies $\mathcal{P} h_f(k) \mathcal{P}^{-1}= -h_f(-k)$. As elucidated further in the next section, it is due to this particle hole symmetry and the existence of two gaps which allow the Majorana condition to hold at quasienergy zero and $\pi$.


}

\section*{Section C: Majorana modes in time periodic systems and Majorana time crystals}

In treating time periodic systems, Floquet formalism \cite{Flo1,Flo2,Flo3,Flo4} is usually employed so that essential information about the systems is encoded in the eigenstates of the one period propagator $\mathcal{U}$ as defined in the main text. The corresponding eigenphases of $\mathcal{U}$ are also referred to as quasienergies (up to a factor $1/T$) as an analogy with eigenenergies in static systems. However, unlike energy, eigenphase of $\mathcal{U}$  is only defined modulo $2\pi$ due to its phase nature. Consequently, eigenphases $\varepsilon=\pm \pi$ are identified as the same.

In the second quantization language, we may define a fermion mode $\Psi_\varepsilon$ associated with each eigenphase $\varepsilon$ of $\mathcal{U}$. By taking $|R\rangle$ as a reference state satisfying $\mathcal{U}|R\rangle =|R\rangle$, we can then construct another eigenstate of $\mathcal{U}$ with eigenphase $\varepsilon$ as $|\varepsilon\rangle = \Psi_\varepsilon^\dagger |R\rangle$. In superconducting systems, the existence of particle-hole symmetry guarantees that associated with a fermion mode $\Psi_\varepsilon$ of eigenphase $\varepsilon$, there exists another fermion mode $\Psi_{-\varepsilon}$ of eigenphase $-\varepsilon$. These two modes are related by

\begin{equation}
\Psi_\varepsilon = \Psi_{-\varepsilon}^\dagger \;.
\label{parhole}
\end{equation}

\n Equation~(\ref{parhole}) implies that $\Psi_0$ and $\Psi_{\pm \pi}$ are Hermitian, which are thus termed Floquet Majorana zero and $\pi$ modes respectively. However, such Hermitian fermion modes must always come in pairs, e.g. $\Psi_0^{(1)}$ and $\Psi_0^{(2)}$, in order to form a complex fermion $\Psi_0^{(c)}=\Psi_0^{(1)}+\mathrm{i}\Psi_0^{(2)}$, since $\mathcal{U}$ can only admit terms of the form $\Psi_0^{(c)\dagger}\Psi_0^{(c)}$. In this sense, Majorana zero and $\pi$ modes are usually also referred to as half-fermions.

The above idea can be readily generalized to define Majorana modes in DTCs. Since discrete time translational symmetry is spontaneously broken, eigenphase $\varepsilon$ is no longer a conserved quantity. However, since generally a time crystal state still exhibits periodicity of period $n\neq 1$ times of that of the Hamiltonian itself, a new reference state $|R'\rangle$ can be defined, which satisfies $\mathcal{U}^n |R'\rangle = |R'\rangle$. New set of fermion modes $\Phi_{\varepsilon_n}$ can then be defined based on the eigenphases $\varepsilon_n$ of $\mathcal{U}^n$, where particle-hole symmetry implies $\Phi_0$ to be Hermitian, which is analogues to Floquet Majorana modes above. DTCs possessing such Majorana modes are what we termed `Majorana time crystals' (MTCs) in the main text.

\section*{Section D: The role of Majorana zero and $\pi$ modes in MTCs}

\begin{figure}
	\begin{center}
		\includegraphics[scale=0.45]{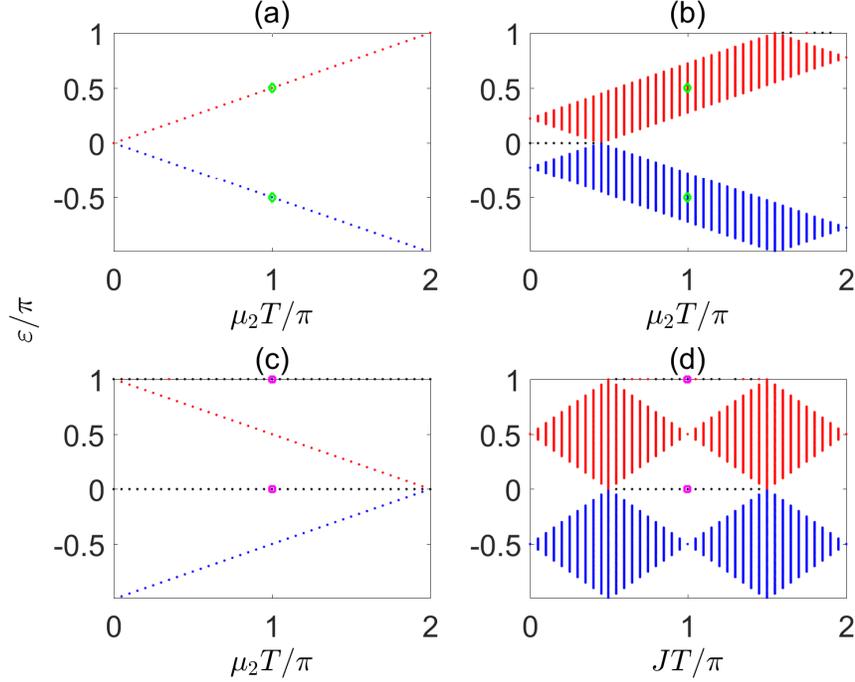}%
	\end{center}
	\caption{(color online). Eigenphase spectrum of $\mathcal{U}$ in the main text as a function of some system parameters obtained under the open boundary conditions.  Red and blue colours are used to distinguish the upper and lower bands, while black dots are used to label Majorana zero and $\pi$ modes. Panel (a)-(c) show the dependence of the eigenphases on $\mu_2$ with (a) $J=\Delta=0$, (b) $JT=\Delta T=0.7 $, and (c) $JT=\Delta T=\pi$. While the existence of bulk DTCs relies on a finite bandwidth in the vicinity of eigenphase $\pm \frac{ \pi}{2}$ (see panel (b)), MTCs can only be formed in the presence of both Majorana zero and $\pi$ modes under the same set of parameter values  (see panel (c)). Panel (d) shows the dependence of the eigenphases on $J=\Delta$ at a fixed $\mu_2 T=\pi$.  Green diamonds and magenta squares show examples of two quasienergies which are separated by $\pi$. Robust co-existence of zero and $\pi$ edge modes can
be clearly seen in panels (c) and (d).}
	\label{supic}
\end{figure}

In order to construct a DTC with period $nT$ ($n\neq 1$), it is necessary that the associated Floquet operator (one period propagator) $\mathcal{U}$ possesses at least $n$ eigenphases, each differing from another by an integer multiple of $\frac{2\pi}{n}$. A state exhibiting $nT$ period can then be constructed as a superposition of such $n$ states with different eigenphases \cite{tcrev}. However, in general there is no guarantee that this periodicity is robust against a small change in the system parameters. For the system considered in the main text, setting $\mu_2 T=\pi$ with $J=\Delta=\mu_1=0$ leads to two highly degenerate eigenphases $\pm \frac{\pi}{2}$, as depicted in Fig.~\ref{supic}(a). However, upon slightly changing the value of $\mu_2$, the two eigenphases move to other values and are no longer separated by $\pi$. As a result, a state formed as a superposition of these two Floquet states will not exhibit a robust periodicity. Hence there is no DTC formed.
For nonzero $J=\Delta<\frac{\pi}{T}$, some degeneracies of the previous type are lifted, and the eigenphase spectrum form two bands of finite bandwidth, as depicted in Fig.~\ref{supic}(b). Owing to this finite bandwidth, two eigenphases at $\pm \frac{ \pi}{2}$ remain in the vicinity of $\mu_2 T=\pi$ [compare the two green diamond marks in panel (a) and (b)]. A superposition of two such states may then exhibit rigid periodicity to some extent, leading to the emergence of DTCs. However, such DTCs are formed by two bulk Floquet states, which are not the main focus of the main text.

To construct Majorana time crystals, we require at least two edge states with eigenphase separation of $\pi$. Majorana zero and $\pi$ modes indeed satisfy this requirement. We thus hope that both Majorana zero and $\pi$ modes exist under the same system parameters. Fortunately, this can be achieved by setting $J=\Delta\approx \frac{\pi}{T}$, as is evident from Figs.~\ref{supic}(c) and (d)  (Magenta squares in both panels). Period-doubling state can then be constructed by creating a superposition of these Majorana zero and $\pi$ modes. Unlike bulk DTCs elucidated above, the rigidity of Majorana time crystals stems from topology, as both Majorana modes are topologically protected and they can only be lifted when the two bulk bands touch each other. Finally, it is worth noting that spontaneous time translational symmetry breaking phenomenon becomes apparent in this context for the following reason. Consider a state initially prepared in a zero edge state at certain system parameter values which only admit Majorana zero modes. Suppose that the parameter values are rapidly changed so that the system now admit both Majorana zero and $\pi$ modes. The state, localized at the edge to begin with, will now naturally become a coherent superposition of all the Floquet eigenstates of the new Hamiltonian with most population being on zero and $\pi$ edge states.  As a result, its periodicity changes from $T$ to $2T$, while the Hamiltonian remains periodic with period $T$.

\section*{Section E: Adiabatic following of Majorana zero and $\pi$ modes in the first two steps of the protocol}

In the Heisenberg picture, the evolution of Majoranas over one period is described as $\gamma(t)=\mathcal{U}^\dagger \gamma(0)\mathcal{U}$. For $\hat{0}$ and $\hat{\pi}$ to be instantaneous Majorana zero and $\pi$ modes during the protocol, it suffices to show that $\mathcal{U}^\dagger \hat{0} \mathcal{U}=\hat{0}$ and $\mathcal{U}^\dagger \hat{\pi} \mathcal{U}=-\hat{\pi}$ for instantaneous values of the system parameters.  Note that $\mathcal{U}$ can be written as a product of two exponentials
\begin{equation}
\mathcal{U}=\exp\left(-\mathrm{i} H_2\right)\times \exp\left(-\mathrm{i} H_1\right),
 \end{equation}
 where $T$ has been set to $2$ for brevity, $H_1$ and $H_2$ are defined in the main text. To simplify our notation, we will focus only on relevant terms in $H_1$ and $H_2$ which contain $\gamma_j^A$ and $\gamma_j^B$ with $j=1,2,3$; we suppress the rest of the terms since they always commute with both $\hat{0}$ and $\hat{\pi}$.

In step 1, we have (using the same parameterization described in the main text)

\begin{eqnarray}
H_1&=&\mathrm{i} \frac{\pi}{2} \left(\sin\phi_1 \gamma^B_2 \gamma^A_1 +\cos\phi_1 \gamma^B_2 \gamma^A_3 +\gamma_1^B \gamma_2^A +\gamma_2^B\gamma_3^A+\gamma_3^B \gamma_4^A\right)\;, \\ H_2&=&-\frac{\mathrm{i}\pi}{4} \left( \gamma_1^A \gamma_1^B+\gamma_2^A\gamma_2^B+\gamma_3^A\gamma_3^B\right)\;,
\end{eqnarray}

\n where $\phi_1$ is adiabatically increased from $0$ to $\pi/2$. According to the main text,

\begin{eqnarray}
\hat{0}&=&\frac{1}{\sqrt{2}}\left\lbrace[\cos(\phi_1)\gamma_1^A-\sin(\phi_1)\gamma_3^A]+ [\cos(\phi_1)\gamma_1^B-\sin(\phi_1)\gamma_3^B]\right\rbrace, \nonumber\\
\hat{\pi}&=&\frac{1}{\sqrt{2}}\left\lbrace[\cos(\phi_1)\gamma_1^A-\sin(\phi_1)\gamma_3^A]- [\cos(\phi_1)\gamma_1^B-\sin(\phi_1)\gamma_3^B]\right\rbrace. \nonumber
\end{eqnarray}

\n It can be easily verified that the first (second) bracket in $\hat{0}$ and $\hat{\pi}$ commute (anti-commute) with $H_1$. By using the identity

\begin{equation}
\exp\left(\theta \gamma_j^A \gamma_k^B\right)=\cos(\theta)+\sin(\theta) \gamma_j^A \gamma_k^B \;,
\label{id}
\end{equation}

\n the first exponential in $\mathcal{U}$ transforms $\hat{0}\rightarrow \hat{\pi}$ and $\hat{\pi}\rightarrow \hat{0}$. Since $H_2$ anti-commute with both $\hat{0}$ and $\hat{\pi}$, it can be shown that applying Eq.~(\ref{id}) on the second exponential in $\mathcal{U}$ leads to transformation $\hat{0}\rightarrow -\hat{\pi}$ and $\hat{\pi} \rightarrow \hat{0}$. Taken together, $\mathcal{U}$ yields the desired result $\hat{0}\rightarrow \hat{0}$ and $\hat{\pi}\rightarrow \hat{0}$.

In the second step,

\begin{eqnarray}
H_1&=&\mathrm{i} \frac{\pi}{2} \left( \gamma^B_2 \gamma^A_1 +\sin(\phi_2) \gamma_3^A \gamma_2^A +\cos(\phi_2)\gamma_1^B \gamma_2^A +\gamma_2^B\gamma_3^A+\gamma_3^B \gamma_4^A\right)\;, \\
H_2&=&-\frac{\mathrm{i}\pi}{4} \left( \gamma_1^A \gamma_1^B+\gamma_2^A\gamma_2^B+\gamma_3^A\gamma_3^B\right)\;,
\end{eqnarray}

\n where $\phi_2$ is adiabatically increased from $0$ to $\pi/2$. According to the main text,
{
\begin{eqnarray}
\hat{0}= [-\cos(\phi_2)\gamma_3^A+\sin(\phi_2)\gamma_1^B]-[\cos(\phi_2)\gamma_3^B+\sin(\phi_2)\gamma_1^A], \nonumber\\
\hat{\pi}=[-\cos(\phi_2)\gamma_3^A+\sin(\phi_2)\gamma_1^B]+[\cos(\phi_2)\gamma_3^B+\sin(\phi_2)\gamma_1^A].\nonumber
\end{eqnarray}
}

\n Using the same approach as before, we consider the action of each exponential in $\mathcal{U}$ separately on $\hat{0}$ and $\hat{\pi}$. Since the first (second) bracket of $\hat{0}$ and $\hat{\pi}$ commute (anti-commute) with $H_1$, it immediately follows that $\exp\left(\mathrm{i} H_1\right)\hat{0}\exp\left(-\mathrm{i} H_1\right)=\hat{\pi}$ and $\exp\left(\mathrm{i} H_1\right)\hat{\pi}\exp\left(-\mathrm{i} H_1\right)=\hat{0}$. Meanwhile, $H_2$ remains anti-commute with both $\hat{0}$ and $\hat{\pi}$, so that $\exp\left(\mathrm{i} H_2\right)\hat{0}\exp\left(-\mathrm{i} H_2\right)=-\hat{\pi}$ and $\exp\left(\mathrm{i} H_2\right)\hat{\pi}\exp\left(-\mathrm{i} H_2\right)=\hat{0}$. Taken together, indeed $\mathcal{U}^\dagger \hat{0} \mathcal{U}=\hat{0}$ and $\mathcal{U}^\dagger \hat{\pi} \mathcal{U}=-\hat{\pi}$.





{
	
\section*{Section F: Tolerance of step 3 against small deformations in the adiabatic parameters}

As elucidated in the main text, step 3 of the protocol corresponds to a rotation between $\gamma_L^A$ and $\gamma_L^B$ according to $\Delta_1= 2\pi f_{3,a}(t)/T$, $J_2=\sqrt{2} \pi\exp\left(\frac{\mathrm{i} \pi}{4}\right)[1-\mathrm{i}f_{3,b}(t)]/T$, and $\Delta_2=\sqrt{2} \pi \exp\left(-\frac{\mathrm{i} \pi}{4}\right) [1+\mathrm{i}f_{3,c}(t)]/T$. In the following, we numerically evaluate the evolution of the Majorana correlation functions under two different choice of the functions $f_{3,l=a,b,c}$ with additional small deformations in the functional forms of the adiabatic parameters as given by $J_2\rightarrow J_2+\delta J_2$, $\Delta_2\rightarrow \Delta_2+\delta \Delta_2$, $\Delta_1\rightarrow \Delta_1+\delta \Delta_1$.

To quantify the precision of our result, we define a normalized fidelity as $\mathcal{F}=\frac{\langle \psi_{f,\mathrm{num}} |\psi_{f,\mathrm{th}} \rangle - \frac{1}{\sqrt{2}}}{1-\frac{1}{\sqrt{2}}}$, where $|\psi_{f,\mathrm{th}}\rangle = \frac{1}{\sqrt{2}}\left(|0\rangle -|1\rangle \right)$, with $|0\rangle$ and $|1\rangle$ being parity eigenstates as defined in the main text, is the expected final state resulted from a perfect braiding operation, and $|\psi_{f,\mathrm{num}}\rangle$ is the final state obtained numerically from the proposed protocol. Note that $\mathcal{F}\leq \langle \psi_{f,\mathrm{num}} |\psi_{f,\mathrm{th}} \rangle$ for any real $\langle \psi_{f,\mathrm{num}} |\psi_{f,\mathrm{th}} \rangle$, with $\mathcal{F}=1$ if $|\psi_{f,\mathrm{th}}\rangle = |\psi_{f,\mathrm{num}}\rangle$, and $\mathcal{F}=0$ if $|\psi_{f,\mathrm{num}} \rangle = |\psi_0\rangle = |0\rangle$. In Fig.~\ref{fidelity}(a), we take $f_{3,a}(t)=f_{3,b}(t)=f_{3,c}(t)=\cos\left[\pi s(t)\right]$ (the same as those used in the main text) with $s(t)$ decreasing from $1$ to $0$. The normalized fidelity is found to be $\mathcal{F}=0.9956$. In Fig.~\ref{fidelity}(b), we take three different functions $f_{3,a}(t)=1-2\sin\left[\pi s(t)/2\right]$, $f_{3,b}=1-2s(t)$, $f_{3,c}=\cos\left[\pi s(t)\right]$, which leads to the normalized fidelity of $\mathcal{F}=0.9949$. These two fidelities are above the typically required threshold level of $\approx 1$\% error for fault-tolerant quantum computation \cite{error,error2}. The fact that such high fidelities are achieved even in the presence of considerable time-dependent perturbation demonstrates certain weak topological protection of our protocol.

\begin{figure}
	\begin{center}
		\includegraphics[scale=0.45]{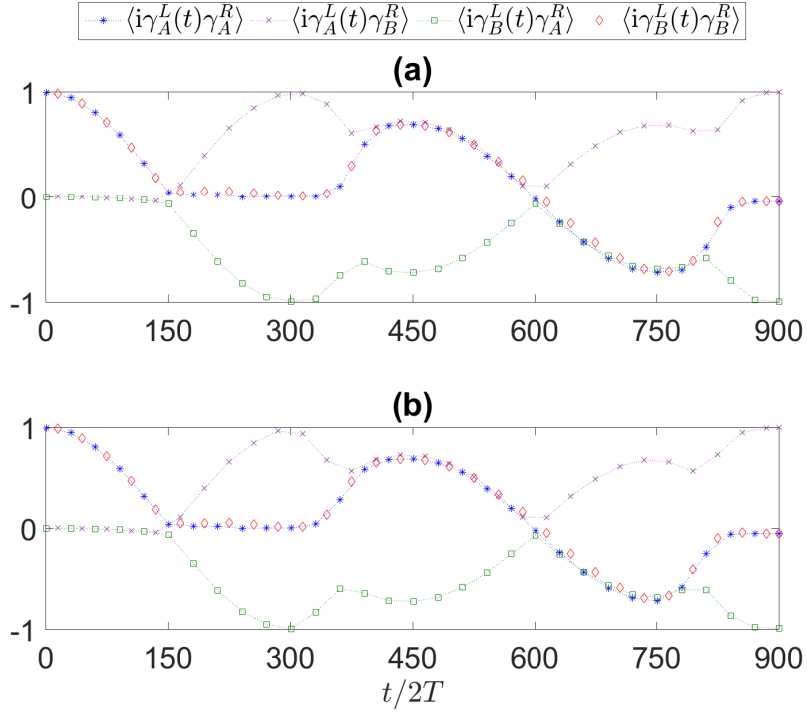}
	\end{center}
	\caption{(color online). Evolution of Majorana correlation functions during the manipulation process in the presence of deformations $\delta J_2 T = (0.23+\mathrm{i} 0.13) \sin\left[\pi s(t)\right]$, $\delta \Delta_2 T=-(0.09+\mathrm{i} 0.15) \sin\left[\pi s(t)\right]$, and $\delta \Delta_1 T= -0.18 \left[s(t)-s(t)^2\right]$, with $s(t)$ increasing from $0$ to $1$, under two different choice of $f_{3,a}$, $f_{3,b}$, $f_{3,c}$ . The other system parameters $JT=3.3$, $\Delta T=3$ $\mu_1 T=0.4$, $\mu_2 T=\pi$, and $N=100$, with each step taking $300$ periods to complete.}
	\label{fidelity}
\end{figure}

\section*{Section G: Experimental consideration of the protocol}

To further verify the feasibility of the proposed protocol in real experiment, it is important to compare the coherence time-scale of the Majorana modes in our system and the typical time required to complete the whole experiment. In the cold atom setup proposed in the main text, after taking into account a variety of mechanisms that induce particle loses, the coherence time-scale can be assumed to be extendable to the order of seconds \cite{opt}. On the other hand, given that the system parameters are typically of the order of tens of kHz \cite{opt}, a single period can last of the order of $0.1$ ms in order to achieve the parameter regime in which Majorana zero and $\pi$ modes exist. As demonstrated by our numerics, each step in the protocol may require around $200$-$400$ number of periods in order to achieve a good precision. This means that to complete a single braiding process, it may take around $1200$-$2400$ number of periods, which is of the order of $0.1$ s. Comparing the two time scales together, it is expected that a number of braiding operations can be applied via our protocol safely before the system loses its effectiveness.

In real experiment, one may also be wary of decoherence. Due to the nonlocal nature of our qubits, local errors induced by decoherence are typically harmless by themselves. However, over time, such errors may accumulate across the whole lattice, leading to a nonlocal error of the form $\mathrm{i}\gamma_L^l \gamma_R^k$, with $l,k=A,B$, which may become dangerous. Therefore, especially if one needs to perform longer time operations, it may be necessary to devise a correction scheme to avoid such an error. Given that Majorana chain is a natural stabilizer code, the most natural error correction scheme will be to measure these (Floquet) stabilizer operators (e.g. $\mathrm{i} \gamma_{j}^A\gamma_{j}^B$ with $j\neq 1,N$) stroboscopically at every period, then perform suitable correction by simply applying the error operators one more time, since local errors $\mathcal{E}_j$ compatible with parity conservation are typically of the form $\mathrm{i} \gamma_j^B \gamma_{j+1}^A$, which squares to identity.  In the proposed cold atom experiment, measuring these stabilizer operators should be feasible as one essentially needs to only measure the parity of each lattice site, while applying corrections can be done by injecting/removing fermions at the infected sites. Although this error correction scheme is rather straightforward, the time required to perform corrections scales with the number of lattice sites, so it may become less efficient for longer wires. As such, it may also be interesting to explore other more sophisticated error correction schemes, such as that introduced in Ref.~\cite{ecor}. However, these are beyond the scope of this work and may be left for potential future studies.

}
\section*{Section H: Extension of the protocol to an array of superconducting wires}

\begin{figure}
	\begin{center}
		\includegraphics[scale=0.45]{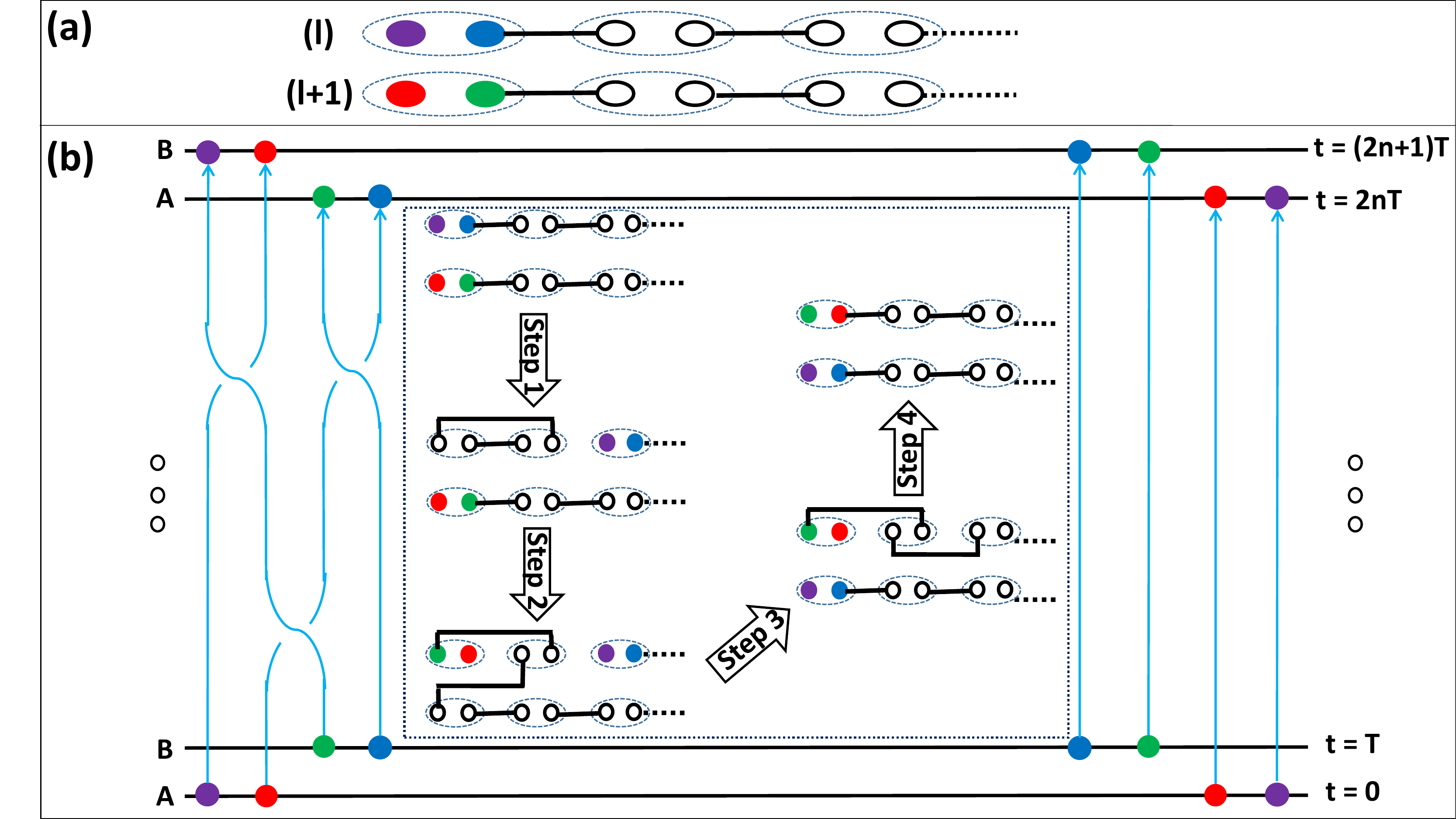}%
	\end{center}
	\caption{(color online). (a) Two adjacent wires labeled $(l)$ and $(l+1)$ in the array and their associated Majorana modes. (b) Schematic of the modified protocol involving Majorana modes from different wires.}
	\label{supic2}
\end{figure}

Since our original setup only allows the implementation of a single qubit in either even or odd parity subspace of the wire,  its manipulation through braiding and its application in quantum computing may at first look quite limited. However, by adding more wires and making some modifications to our protocol, it is possible to encode and manipulate more qubits, thus unleashing the full power of MTCs to perform more complex quantum computation.

To elaborate on this idea, consider an array of superconducting wires, each described by Hamiltonian $H^{(l)}(t)$ as defined in the main text, where an index $l$ has been introduced to mark different wires in the array. In particular, the index $l$ is given to all system parameters, namely $J_j^{(l)}$, $\Delta_j^{(l)}$, $\mu_1^{(l)}$, and $\mu_2^{(l)}$ which to some extents allow each wire not to be fully identical. We will focus our attention on two adjacent wires $l$ and $l+1$ as shown in Fig.~\ref{supic2}(a). Upon turning on the periodic driving, the Majorana zero modes at each wire acquire period-doubling behaviour, which leads to four pairs of Majoranas in space-time lattice. In particular, we may denote the four Majoranas localized at the left edge as $\gamma_A^{(l)}$, $\gamma_A^{(l+1)}$, $\gamma_B^{(l)}$, and $\gamma_B^{(l+1)}$, where the subscript and superscript indices denote the time lattice and wire array respectively.

The modified protocol consists of four steps, which are summarized in the inset of Fig.~\ref{supic2}(b). In the first step, we modify $J_1^{(l)}$, $J_2^{(l)}$, $\Delta_1^{(l)}$ and $\Delta_2^{(l)}$ according to step 1 in our original protocol, while the rest of parameters are kept constant. This results in $\gamma_A^{(l)}$ and $\gamma_B^{(l)}$ to localize around the third site of the $l$th-wire. In the second step, we introduce hopping and pairing between the first site of the $(l+1)$th-wire and the second site of the $l$th-wire, while at the same time varying $J_1^{(l)}$ and $\Delta_1^{(l)}$ according to step 2 in our original protocol and keeping the other parameters fixed. This moves $\gamma_A^{(l+1)}$ and $\gamma_B^{(l+1)}$ to the first site of the $l$th-wire. In the third step, we reduce the hopping and pairing strength between the first site of the $(l+1)$th-wire and the second site of the $l$th-wire adiabatically to zero, while at the same time varying $J_2^{(l)}$ and $\Delta_2^{(l)}$ according to step 2 in our original protocol and keeping the other parameters constant. This moves $\gamma_A^{(l)}$ and $\gamma_B^{(l)}$ to the first site of the $(l+1)$th-wire. Finally, in the last step, we vary $J_1^{(l)}$, $J_2^{(l)}$, $\Delta_1^{(l)}$ and $\Delta_2^{(l)}$ according to step 3 and 4 in our original protocol to return the Hamiltonian to its original configuration. The net outcome of the aforementioned protocol simulates a successive braiding between different pairs of Majoranas as $\gamma_A^{(l+1)}\rightarrow \gamma_B^{(l+1)}\rightarrow -\gamma_B^{(l)}$,  $\gamma_B^{(l+1)}\rightarrow -\gamma_A^{(l+1)}\rightarrow \gamma_A^{(l)}$, $\gamma_A^{(l)}\rightarrow \gamma_A^{(l+1)}$, and $\gamma_B^{(l)}\rightarrow \gamma_B^{(l+1)}$, which is described schematically in Fig.~\ref{supic2}(b).  This makes it clear that we can scale up the MTCs described in the main text to implement multi-gate operations.



\begin{thebibliography}{99}
		\bibitem{Wilczek} F.~Wilczek, \prl ~{\bf 109}, 160401 (2012).
		\bibitem{no} H.~Watanabe and M.~Oshikawa, \prl ~{\bf 114}, 251603 (2015).
		\bibitem{Sacha} K.~Sacha, \pra ~{\bf 91}, 033617 (2015).
		\bibitem{the1} D.~V.~Else, B.~Bauer, and C.~Neyak, Phys. Rev. Lett. ~{\bf 117}, 090402 (2016).
		\bibitem{the2} N.~Y.~Yao, A.~C.~Potter, I.-D.~Potirniche, and A.~Vishwanath, \prl ~{\bf 118}, 030401 (2017).
		\bibitem{the3} W.~W.~Ho, S.~Choi, M.~D.~Lukin, and D.~A.~Abanin, \prl ~{\bf 119}, 010602 (2017).
		\bibitem{the4} D.~V.~Else, B.~Bauer, and C.~Neyak, Phys. Rev. X ~{\bf 7}, 011026 (2017).
		\bibitem{the5} B.~Huang, Y.-H.~Wu, and W.~V.~Liu, arXiv:1703.04663v1.
		\bibitem{the6} A.~Russomanno, F.~Lemini, M.~Dalmonte, and R.~Fazio, \prb ~{\bf 95}, 214307 (2017).
		{  \bibitem{the7} A.~Russomanno, B. Friedman, and E.~G.~Dalla Torre, \prb ~{\bf 96}, 045422 (2017). }
		\bibitem{exp1} J.~Zhang, P.~W.~Hess, A.~Kyprianidis, P.~Becker, A.~Lee, J.~Smith, G.~Pagano, I.-D.~Potirniche, A.~C.~Potter, A.~Vishwanath, N.~Y.~Yao, and C.~Monroe, Nature ~{\bf 543}, 217 (2017).
		\bibitem{exp2} S.~Choi, J.~Choi, R.~Landig, G.~Kucsko, H.~Zhou, J.~Isoya, F.~Jelezko,
		S.~Onoda, H.~Sumiya, V.~Khemani, C.~v.~Keyserlingk, N.~Y.~Yao, E.~Demler, and
		M.~D.~Lukin, Nature ~{\bf 543}, 221 (2017).
		\bibitem{tcrev} K.~Sacha and J.~Zakrewski, Rep. Prog. Phys. ~{\bf 81}, 016401 (2017). 
		\bibitem{tqc} C.~Nayak, S.~H.~Simon, A.~Stern, M.~Freedman, and S.~Das~Sarma, Rev. Mod. Phys. ~{\bf 80}, 1083 (2008). 
		\bibitem{tqc2} V.~Lahtinen and J.~K.~Pachos, SciPost Phys. ~{\bf 3}, 021 (2017). 
		\bibitem{ising} A.~Kitaev, Ann.~Phys. ~{\bf 321}, 2 (2006). 
		\bibitem{ising2} A.~Ahlbrecht, L.~S.~Georgiev, and R.~F.~Werner, \pra ~{\bf 79}, 032311 (2009). 
		\bibitem{ising3} G.~Moore and N.~Read, Nucl. Phys. B ~{\bf 360}, 362 (1991). 
		\bibitem{fib} S.~Trebst, M.~Troyer, Z.~Wang, and A.~W.~W.~Ludwig, Prog. Theor. Phys. Supp. ~{\bf 176}, 384 (2008). 
		\bibitem{fib2} R.~S.~K.~Mong, D.~J.~Clarke, J.~Alicea, N.~H.~Lindner, P.~Fendley, C.~Nayak, Y.~Oreg, A.~Stern, E.~Berg, K.~Shtengel, and M.~P.~A Fisher, Phys. Rev. X ~{\bf 4}, 011036 (2014). 
		\bibitem{Ivanov} D.~A.~Ivanov, Phys. Rev. Lett. ~{\bf 86}, 268 (2001). 
		\bibitem{Kit} A.~Y.~Kitaev, Phys. Usp ~{\bf 44}, 131 (2001). 
		\bibitem{mich} M.~Stone and S.-B~Chung, \prb ~{\bf 73}, 014505 (2006). 
		
		\bibitem{map} P. Fendley, J. Stat. Mech. ~{\bf 11}, 20 (2012). 
		\bibitem{DLoss} F.~L.~Pedrocchi, S.~Chesi, S.~Gangadharaiah, and D.~Loss, \prb ~{\bf 86}, 205412 (2012). 
	\bibitem{Ychen} Y.-C.~He and Y.~Chen, \prb ~{\bf 88}, 180402(R) (2013). 
\bibitem{opt} L.~Jiang, T.~Kitagawa, J.~Alicea, A.~R.~Akhmerov, D.~Pekker, G.~Refael, J.~I.~Cirac, E.~Demler, M.~D.~Lukin, and P.~Zoller, \prl ~{\bf 106}, 220402 (2011). 
    \bibitem{kk3} Q.-J.~Tong, J.-H.~An, J.~B.~Gong, H.-G.~Luo, and C.~H.~Oh, \prb ~{\bf 87}, 201109(R) (2013). 
     \bibitem{M3} D.~E.~Liu, A.~Levchenko, and H.~U.~Baranger, \prl~{\bf 111}, 047002 (2013).
        \bibitem{Tjun1} J.~Alicea, Y.~Oreg, G.~Refael, F.~von~Oppen, and M.~P.~A.~Fisher, Nat. Phys. ~{\bf 7}, 412 (2011). 
		\bibitem{Tjun2} B.~van~Heck, A.~R.~Akhmerov, F.~Hassler, M.~Burrello, and C.~W.~J.~Beenakker, New J. Phys. ~{\bf 14}, 035019 (2012). 
	
		\bibitem{wire} C.~V.~Kraus, P.~Zoller, and M.~A.~Baranov, \prl ~{\bf 111}, 203001 (2013). 
		\bibitem{Tjun3} T.~Karzig, F.~Pientka, G.~Refael, and F.~von~Oppen, \prb ~{\bf 91}, 201102 (2015). 
		\bibitem{Tjun4} P.~Gorantla and R.~Sensarma, arXiv:1712.00453. 
		\bibitem{magic} T.~Karzig, Y.~Oreg, G.~Refael, and M.~H.~Freedman, Phys. Rev. X ~{\bf 6}, 031019 (2016). 
		\bibitem{magic2} S.~Bravyi and A.~Kitaev, \pra ~{\bf 71}, 022316 (2005). 
		\bibitem{magic3} S.~Bravyi, \pra ~{\bf 73}, 042313 (2006). 
		\bibitem{magic4} M.~Freedman, C.~Nayak, and K.~Walker, \prb ~{\bf 73}, 245307 (2006).
		\bibitem{magic5} P.~Bonderson, D.~J.~Clarke, C.~Nayak, and K.~Shtengel, \prl ~{\bf 104}, 180505 (2010). 
\bibitem{Supp} See Supplemental Material, which includes Refs.~\cite{Flo1,Flo2,Flo3,Flo4,error,error2,ecor}
\bibitem{Flo1} J.~H.~Shirley, Phys. Rev. ~{\bf 138}, B979 (1965). 
\bibitem{Flo2} H.~Sambe, \pra ~{\bf 7}, 2203 (1973). 
\bibitem{Flo3} T.~Oka and H.~Aoki, \prb ~{\bf 79}, 081406 (2009). 
\bibitem{Flo4} N.~H.~Lindner, G.~Refael, and V.~Galitski, Nat. Phys. ~{\bf 7}, 490 (2011). 
\bibitem{error} E.~Knill, Nature {\bf 434}, 39 (2005). 
\bibitem{error2} C.~J.~Ballance, T.~P.~Harty, N.~M.~Linke, M.~A.~Sepiol, and D.~M.~Lucas, \prl ~{\bf 117},060504 (2016). 
	\bibitem{ecor} N.~Lang and H.~P.~B\"{u}chler, SciPost Phys. ~{\bf 4}, 007 (2018).

    \bibitem{Derek1} D.~Y.~H.~Ho and J.~B.~Gong, \prb~{\bf 90}, 195419 (2014).
    \bibitem{Zhou1} L.~W.~Zhou, H.~L.~Wang, D.~Y.~H.~Ho and J.~B.~Gong, EPJB~{\bf 87}, 204 (2014).
    \bibitem{R1}R.~W.~Bomantara, G.~N.~Raghava, L.~W.~Zhou, and J.~B.~Gong, \pre~{\bf 93}, 022209 (2016).
		\bibitem{kk1} M.~N.~Chen, F.~Mei, W.~Shu, H.-Q.~Wang, S.-L.~Zhu, L.~Sheng, and D.~Y.~Xing, J. Phys.: Condens. Matter ~{\bf 29}, 035601 (2016). 
		\bibitem{kk2} H.-Q.~Wang, M.~N.~Chen, R.~W.~Bomantara, J.~B.~Gong, and D.~Y.~Xing, \prb ~{\bf 95}, 075136 (2017). 
		
		\bibitem{tof} C.~V.~Kraus, S.~Diehl, M.~A.~Baranov, and P.~Zoller, New J. Phys. {\bf 14}, 113036 (2012).
		\bibitem{read} J.~F.~Sherson, C.~Weitenberg, M.~Endres, M.~Cheneau, I.~Bloch, and S.~Kuhr, Nature ~{\bf 467}, 68 (2010).
		\bibitem{note} 
 This exchange $\gamma_L^A \leftrightarrow \gamma_L^B$ is trivial as compared with the braiding exchange $\gamma_L^A\rightarrow \gamma_L^B$ and $\gamma_L^B\rightarrow -\gamma_L^A$ (i.e., an extra negative sign is necessary for a braiding process).
	\end{thebibliography}
\end{document}